\documentclass[aps,prl,twocolumn,showpacs,showkeys,superscriptaddress]{revtex4-1}  
\usepackage{graphicx,color}  
\usepackage{float}        
\usepackage{amsmath}
\usepackage{amssymb}   
\usepackage{xfrac}   
\usepackage{mathptmx}
\usepackage{comment}
\usepackage{tikzsymbols}
\usepackage{tikz}
\usepackage[utf8]{inputenc}
\newcommand{\wechselc}[6][1]{\scalebox{#1}{
		\begin{tikzpicture}[baseline=-\the\dimexpr\fontdimen22\textfont2\relax]
		\node (1) at (-.8,.35) {\(#2\)};
		\node (2) at (-.8,-.35) {\(#3\)};
		\node (3) at (.8,0.45) {\(#4\)};
		\node (4) at (.8,0) {\(#5\)};
		\node (5) at (.8,-.45) {\(#6\)};
		\node[shape=circle,fill=black,inner sep=2pt] (6) at (0,0) {};
		\draw (1) to (6) (2) to (6) (3) to (6) (4) to (6) (5) to (6);
		\end{tikzpicture}}}
\newcommand{\wechselcc}[6][1]{\scalebox{#1}{
		\begin{tikzpicture}[baseline=-\the\dimexpr\fontdimen22\textfont2\relax]
		\node (1) at (-.8,.45) {\(#2\)};
		\node (2) at (-.8,.0) {\(#3\)};
		\node (3) at (-.8,-0.45) {\(#4\)};
		\node (4) at (.8,.35) {\(#5\)};
		\node (5) at (.8,-.35) {\(#6\)};
		\node[shape=circle,fill=black,inner sep=2pt] (6) at (0,0) {};
		\draw (1) to (6) (2) to (6) (3) to (6) (4) to (6) (5) to (6);
		\end{tikzpicture}}}
\newcommand{\wechselccc}[5][1]{\scalebox{#1}{
		\begin{tikzpicture}[baseline=-\the\dimexpr\fontdimen22\textfont2\relax]
		\node (1) at (-.8,.35) {\(#2\)};
		\node (2) at (-.8,-.35) {\(#3\)};
		\node (3) at (.8,.35) {\(#4\)};
		\node (4) at (.8,-.35) {\(#5\)};
		\node[shape=circle,fill=black,inner sep=2pt] (5) at (0,0) {};
		\draw (1) to (5) (2) to (5) (3) to (5) (4) to (5);
		\end{tikzpicture}}}

\renewcommand{\c}{\text{c}}
\newcommand{\dd}{\text{d}}
\newcommand{\s}{\text{s}}

\begin{document}
	
\title{Kinetic approach to a relativistic BEC with inelastic processes}

\author{Richard Lenkiewicz}
\email{lenkiewicz@th.physik.uni-frankfurt.de} 
\affiliation{Institut f{\"u}r Theoretische Physik, Goethe-Universit{\"a}t
  Frankfurt, Max-von-Laue-Stra{\ss}e 1, D-60438 Frankfurt am Main, Germany}
\author{Alex Meistrenko}
\affiliation{Institut f{\"u}r Theoretische Physik, Goethe-Universit{\"a}t
  Frankfurt, Max-von-Laue-Stra{\ss}e 1, D-60438 Frankfurt am Main, Germany}
\author{Hendrik van Hees}
\email{hees@th.physik.uni-frankfurt.de}
\affiliation{Institut f{\"u}r Theoretische Physik, Goethe-Universit{\"a}t
  Frankfurt, Max-von-Laue-Stra{\ss}e 1, D-60438 Frankfurt am Main, Germany}
\author{Kai Zhou}
\affiliation{Frankfurt Institute for Advanced Studies,
  Ruth-Moufang-Stra{\ss}e 1, D-60438 Frankfurt
  am Main, Germany}
\affiliation{Institut f{\"u}r Theoretische Physik, Goethe-Universit{\"a}t
  Frankfurt, Max-von-Laue-Stra{\ss}e 1, D-60438 Frankfurt am Main, Germany}
\author{Zhe Xu}
\affiliation{Department of Physics, Tsinghua University and
  Collaborative Innovation Center of Quantum Matter, Beijing 100084,
  China}
\author{Carsten Greiner}
\affiliation{Institut f{\"u}r Theoretische Physik, Goethe-Universit{\"a}t
  Frankfurt, Max-von-Laue-Stra{\ss}e 1, D-60438 Frankfurt am Main, Germany}

\date{June 18, 2019}

\begin{abstract}
  The phenomenon of Bose-Einstein condensation is investigated in the
  context of the color-glass-condensate description of the initial state
  of ultrarelativistic heavy-ion collisions. For the first time, in this
  paper we study the influence of particle-number changing
  $2 \leftrightarrow 3$ processes on the transient formation of a
  Bose-Einstein condensate within an isotropic system of scalar bosons
  by including $2 \leftrightarrow 3$ interactions of massive bosons with
  constant and isotropic cross sections, following a Boltzmann
  equation. The one-particle distribution function is decomposed in a
  condensate part and a nonzero momentum part of excited modes, leading
  to coupled integro-differential equations for the time evolution of
  the condensate and phase-space distribution function, which are then
  solved numerically. Our simulations converge to the expected
  equilibrium state, and only for $\sigma_{23}/\sigma_{22} \ll 1$ we
  find that a Bose-Einstein condensate emerges and decays within a
  finite lifetime in contrast to the case where only binary scattering
  processes are taken into account, and the condensate is stable due to
  particle-number conservation. Our calculations demonstrate that
    Bose-Einstein condensates in the very early stage of heavy-ion
    collisions are highly unlikely, if inelastic collisions are
    significantly participating in the dynamical gluonic evolution.
\end{abstract}
\keywords{Ultrarelativistic heavy ion collisions; Boltzmann equation; Bose Einstein
  Condensation; Early stage of heavy-ion collisions; Non-equilibrium dynamics for highly occupied systems}
\maketitle
\section{introduction}
	
A deconfined system of quarks and gluons, under extreme conditions of
high temperatures and high densities, can be produced and explored in
experiments of ultrarelativistic heavy-ion collisions. The experimental
observables like elliptic-flow measurements strongly suggest an early
collective-fluid behavior of a medium close to local thermal
equilibrium. However, the description of the pre-thermalization dynamics
of the initial off-equilibrium many-body system produced in heavy-ion
collisions is still an outstanding problem.
	
The early stage of heavy-ion collisions is well described within the
color-glass-condensate (CGC) effective field
theory~\cite{McLerran:2008es,Gelis:2010nm}, where the heavy nuclei
behave as very dense gluon system with high energetic colored partons
acting as sources of soft dynamical gluon fields. In this picture,
during the collision the hard partons traverse each other while the
highly occupied soft gluon fields interact via non-Abelian interactions
resulting in the creation of longitudinal chromo-electric and -magnetic
fields, which leads to the so-called
Glasma~\cite{Weigert:2005us,Lappi:2006fp,Gelis:2010nm,Gelfand:2016yho}
state of high gluon density, which runs through a very short
isotropization stage~\cite{Kurkela:2015qoa,Gelis:2013rba}. Given the
high particle density which is parametrically larger compared to the
thermal-equilibrium value, the system would possess a strongly
interacting nature due to coherently enhanced scattering even though the
coupling is weak. Thus the possible formation of an off-equilibrium
Bose-Einstein condensate (BEC) has drawn stronger attention in recent
years~\cite{Blaizot:2011xf,Blaizot:2012qd}. Similar issues about
off-equilibrium BEC formations arise also in the context of early
universe reheating after inflation
\cite{Prokopec:1997inflaton,Pantel:2012trap_anharmonicity} and in
systems of cold atoms
\cite{Pantel:2012trap_anharmonicity,Lacaze:2001_dynamical_BEC}.

The formation of a BEC is a fundamental consequence of quantum
statistics, where above a certain critical density or below a certain
critical temperature any more added bosons must occupy the ground state
coherently. The condensation dynamics, especially far from equilibrium,
is an interesting issue but still under debate. Many studies have been
performed to understand the nonequilibrium dynamics of BECs formation
within either a kinetic approach or classical field theory, if solely
elastic processes are
incorporated~\cite{Blaizot:2011xf,Berges:2012us,Blaizot:2013lga,Xu:2014ega,MHZC16,EGJGW15,PhysRevD.96.014020}.

Inelastic scattering may qualitatively change the picture, allowing only
for the formation of a transient BEC. In \cite{Huang:2013lia} it is
found that inelastic collisions will speed up the thermalization in the
infrared regime and may catalyze a faster onset of a BEC. The following
study \cite{Blaizot:2016iir} suggested a complete hindrance of BEC
formation for massless gluons at vanishing momentum. Within the
description of a nonequilibrium massive bosonic O($N$) theory applying
the 2PI formalism of real-time Schwinger-Keldysh quantum field theory it
has been recently shown that the formation of a BEC is potentially
prevented by particle-number changing processes
\cite{Tsutsui:2017uzd}. However, a concrete kinetic simulation for a
possible transient BEC has not been included in these studies.
	
So far no kinetic description has been elaborated to describe the
expected transient formation and decay of a BEC, initially possible in
an off-equilibrium system, including both elastic and inelastic
processes. This paper addresses the dynamics of the condensation and
thermalization of massive bosons. For this a coupled set of Boltzmann
kinetic equations for a transient BEC and a phase-space distribution
function is formulated and includes $2 \rightarrow 2$ and
particle-number changing $2 \leftrightarrow 3$ reactions.
\begin{figure*}[t!]
  \centering
  \includegraphics[width=1.07\textwidth]{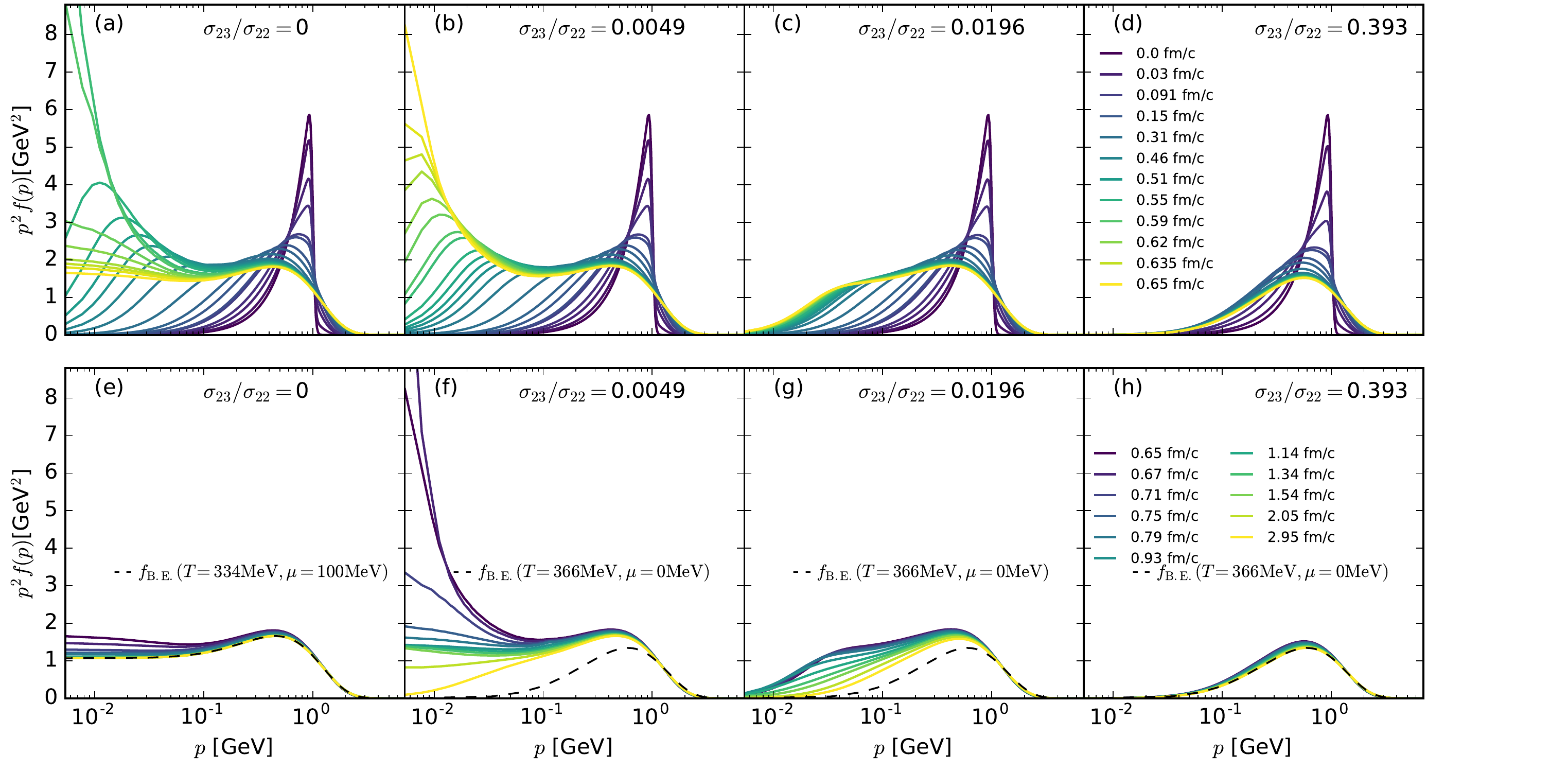}
  \caption{Differential fraction of the isotropic
    particle density with respect to the momentum and for four different
    regimes of the cross section ratio, only binary scattering (a+e),
    elastically dominated scattering (b+f), balanced scattering (c+g) and
    inelastically dominated (d+h), with $m=100$ MeV at $f_0=0.45$ at
    various times. The rows separate two sequential time periods. In (a+e)
    and (b+f) a condensate is present, but in (f) it decays. The black dashed
    lines (bottom row) depicts the individual expected equilibrium
    states.\label{fig:MainResult1_6plots}}
\end{figure*}
	
\section{Kinetic equations}
	
In this work, we focus on an isotropic and homogeneous system. If the
evolution is dominated by two- and three-body interactions, the
corresponding Boltzmann equation for a phase-space distribution function
$f(\vec{p})=d_g \dd N /(2 \pi)^3 \dd^3 x \, \dd^3 p$, where
$d_g=16$ is the gluon degeneracy factor, taking two spin and eight color
states into account, reads \cite{Weinstock:2005jw}
\begin{widetext}
  \begin{equation}
    \begin{split}
      \label{1}
      \dot{f}_1 = & -\frac{1}{2}\frac{1}{2E_1} \int
      \text{d}^9(\vec{2}\vec{3}\vec{4})\delta^{(4)}(1+2-3-4) \left
        \{|M_{12\leftrightarrow 34}|^2
        f_1f_2\left(\frac{f_3}{d_g}+1\right) \left (\frac{f_4}{d_g}+1
        \right) -|M_{12\leftrightarrow 34}|^2f_3f_4
        \left(\frac{f_1}{d_g}+1 \right)
        \left(\frac{f_2}{d_g}+1 \right )\right \} \\
      & -\frac{1}{3!}\frac{1}{2E_1} \int
      \text{d}^{12}(\vec{2}\vec{3}\vec{4}\vec{5})\delta^{(4)}(1+2-3-4-5)
      \Bigg \{|M_{12 \rightarrow 345}|^2 f_1f_2 \left(\frac{f_3}{d_g}+1
      \right)
      \left(\frac{f_4}{d_g}+1 \right) \left(\frac{f_5}{d_g}+1 \right) \\
      &-|M_{345 \rightarrow 12}|^2 f_3 f_4 f_5 \left(\frac{f_1}{d_g}+1
      \right) \left(\frac{f_2}{d_g}+1 \right) \Bigg \}
      -\frac{1}{2!2!}\frac{1}{2E_1} \int
      \text{d}^{12}(\vec{2}\vec{3}\vec{4}\vec{5})\delta^{(4)}(1+2+3-4-5)\\
      & \left \{|M_{123 \rightarrow 45}|^2 f_1f_2f_3
        \left(\frac{f_4}{d_g}+1 \right) \left(\frac{f_5}{d_g}+1 \right)
        -|M_{45 \rightarrow 123}|^2f_4f_5 \left(\frac{f_1}{d_g}+1
        \right) \left(\frac{f_2}{d_g}+1 \right) \left(\frac{f_3}{d_g}+1
        \right) \right \}.
    \end{split}
  \end{equation}
\end{widetext}

Thereby, the indices refer to the momentum phase space of the
participating particles
($\text{d}\vec{i}:=\sfrac{\text{d}^3\vec{p}_i}{2(2\pi)^3E_i}$) with
$E_i=\sqrt{p_i^2+m^2}$. Consequently, $f_i$ denotes the corresponding
one-particle distribution function $f(t,\vec{p}_i)$. The collision
integrals take into account quantum statistics via Bose enhancement
factors $(f_i/d_g+1)$, leading to the correct long-time equilibrium
solution for bosons.  The matrix elements $|M|^2$ are taken as isotropic
with a constant cross section \cite{El:2012cr,XG05}
\begin{equation}
\begin{split}
  |M_{2\leftrightarrow 2}|^2 &= 32\pi s\sigma_{22}, \\
  |M_{2\rightarrow 3}|^2 &=192\pi^3\sigma_{23},\\
  |M_{3\rightarrow 2}|^2 &=\frac{1}{d_g}|M_{2\rightarrow 3}|^2 
\end{split}
\end{equation}
with $s=(P_1+P_2)^2$ denoting the center-momentum energy squared. Here,
we point out that the interesting quantity for the simulations is the
ratio of the elastic and inelastic cross sections,
$\sigma_{23}/\sigma_{22}$, determining the dominating processes leading
to full equilibration.  Energy and particle densities are respectively
given by
\begin{equation*}
  \epsilon_\text{part}(t)=\int\frac{\text{d}^3\vec{p}}{(2\pi)^3}Ef(p) %
  \hspace*{0.2cm} \text{and} \hspace*{0.2cm}n_\text{part}(t)=\int\frac{\text{d}^3\vec{p}}{(2\pi)^3}f(p).
\end{equation*} 
The general argument for the emergence of a BEC is that if in the case
of the existence of conserved number of bosons the chemical potential
converges to the mass, the distribution can no longer accommodate the
particles in the IR regime ($p\ll m$) although
\begin{equation}
  \lim_{\mu\rightarrow m}f_{\text{B.E.}}(p\ll m)=\lim_{\mu\rightarrow m}\frac{d_g}{\exp(\frac{m-\mu}{T})-1}=\infty.
\end{equation}
In this case, a special treatment is necessary for the zero mode, by
decomposing $f(|\vec{p}|)$ in a continuumlike part $f(|\vec{p}|>0)$ for
the higher modes and a discrete part
$(2\pi)^3n_\c(t)\delta^{(3)}(\vec{p})$ for the zero mode
\cite{PhysRevLett.74.3093,Xu:2014ega,MHZC16,PhysRevD.96.014020}.
	
Given any initial nonequilibrium configuration of the gluon system, one
can always determine via the conservation laws if condensation has to be
expected in the equilibrium limit by solving
\begin{equation}\label{conservation1}
  \epsilon_\text{init}=\epsilon_\text{eq}(T,\mu)\hspace*{0.2cm} \text{and} \hspace*{0.2cm} 
  n_\text{init}=n_\text{eq}(T,\mu)
\end{equation}
and if one encounters $\mu > m$ as solution of
Eqs.~\eqref{conservation1}
\begin{equation}
  \epsilon_\text{init}=\epsilon_\text{eq}(T,\mu=m)+\epsilon_\c \hspace*{0.2cm}\text{and} \hspace*{0.2cm} 
  n_\text{init}=n_\text{eq}(T,\mu=m)+n_\c
\end{equation}
where $\epsilon_\c$ and $n_\c$ are the energy and particle density of
the condensate. Those considerations only apply for number conserving
scattering processes ($2\leftrightarrow2$).  However, if one introduces
particle-number changing $2\leftrightarrow3$ scattering processes, this
argument breaks down for massive particles, because in thermal
equilibrium necessarily $\mu=0$, implying that a stable condensate can
not exist.

By inserting the ansatz
$f(p)=f_{|\vec{p}|>0}+(2\pi)^3n_\c(t)\delta^{(3)}(\vec{p})$ into Eq.\
(\ref{1}), we obtain the following evolution equation for the nonzero
momentum modes,
\begin{equation}
  \begin{split}
    \label{eq:diagrams1}
    \dot{f}_{1} = &\wechselccc[0.8]{g}{g}{g}{g} +  \wechselccc[0.8]{g}{\c}{g}{g} + 2\wechselccc[0.8]{g}{g}{\c}{g}\\
    &+\wechselc[0.8]{g}{g}{g}{g}{g}+3\wechselc[0.8]{g}{g}{\c}{g}{g}+\wechselc[0.8]{g}{\c}{g}{g}{g}\\
    &+3\wechselc[0.8]{g}{g}{\c}{\c}{g} +\wechselc[0.8]{g}{g}{\c}{\c}{\c} +\wechselcc[0.8]{g}{g}{g}{g}{g}\\
    &+2\wechselcc[0.8]{g}{g}{g}{\c}{g}+2\wechselcc[0.8]{g}{g}{\c}{g}{g}+\wechselcc[0.8]{g}{\c}{\c}{g}{g}
  \end{split}
\end{equation}
and a rate equation for the condensate density,
\begin{equation}
  \begin{split}
    \label{eq:diagrams2} 
    \dot{n}_\c =& \wechselccc[0.8]{\c}{g}{g}{g} +
    \wechselc[0.8]{\c}{g}{g}{g}{g} + \wechselcc[0.8]{\c}{g}{g}{g}{g} \\
    &+ 2 \wechselcc[0.8]{\c}{\c}{g}{g}{g}
    +\wechselcc[0.8]{\c}{\c}{\c}{g}{g}.
  \end{split}
\end{equation}
Every possible diagrammatic contribution displayed in Eqs.\
(\ref{eq:diagrams1}) and (\ref{eq:diagrams2}), is related to a specific
collision integral, with $\c$ (condensate) and $g$ (gluon) denoting the
participants of the scattering process. The numerical factors relate to
the combinatorial weight of the diagrams. Details are straightforward
but utmost lengthy.
	
For the isotropic case the scattering angles can be integrated out
analytically, leaving us with one-, two- and three-dimensional collision
integrals, which can be solved numerically. The distribution function is
discretized, with the grid becoming finer in the low-momentum
region. For the differential equations we employ an efficient high-order
adaptive Runge-Kutta method (Cash-Karp)
\cite{Cash:1990:VOR:79505.79507}, while the collision integrals are
treated with two different integration methods. For the one- and
two-dimensional integrals we use the simple Simpson quadrature method,
and for the three dimensional integrals we employ the Vegas Monte Carlo
integration routine from \cite{1978JCoPh..27..192L}.
	
\section{Initial Condition}
	
\begin{figure}
  \centering
  \includegraphics[width=0.45\textwidth]{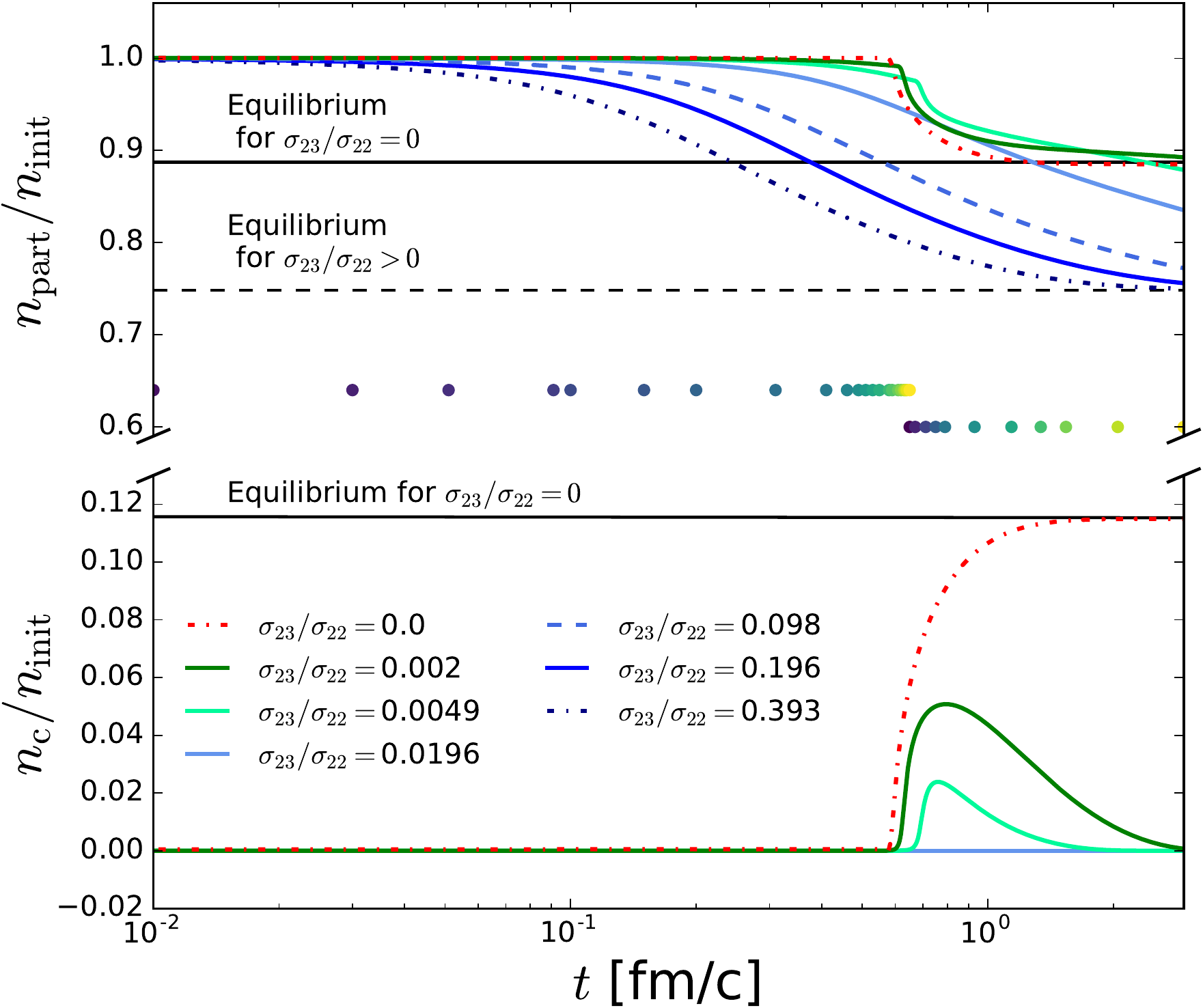}
  \caption{Time evolution of normalized particle (upper
    panel) and condensate (lower panel) densities for particles with
    $m=100$ MeV and $f_0=0.45$. The bluish curves show runs without
    onset of condensation $\mu_\text{eff}<m$.  The scattered dots refer
    to the timestamps of Fig.\ref{fig:MainResult1_6plots}, and the black
    lines mark the individual expected equilibrium
    values.\label{fig:MainResult2}}
\end{figure}
In the context of the CGC framework, the two most relevant quantities
are given by the saturation scale $Q_\s$ and the coupling strength
$\alpha_\s$, which determines the initial population density
$\propto 1/\alpha_\s$ of the initial state. As an initial nonequilibrium
isotropic profile for gluons, formed at time scales of approximately
$1/Q_{\text{s}}$, one usually considers a step function of the form
$f_{\text{init}}(p)=d_gf_0\Theta(1-p/Q_\s)$, whereby
$f_0\sim1/\alpha_\s$
\cite{Blaizot:2011xf,Xu:2014ega,MHZC16,PhysRevD.96.014020}. However, we
use a similar function with a smooth tail around $p\approx Q_s$. Fixing
$Q_s$ at 1 GeV, the only free parameter left is $f_0$, the step
height. Various studies have shown that from this initialization two
scenarios can be observed, if equilibration dynamics are dominated by
binary scattering. The first is the underpopulated case ($f_0<f_\c$)
where the chemical potential never reaches the mass and the second as
the overpopulated case ($f_0>f_\c$), where $\mu=m$ and consequently a
BEC must emerge.  Our investigation is focused on particles with masses
$m=100 (300,500) \, \text{MeV}$ and a cross section of
$\sigma_{22}=1 \, \text{mb}$. These values are close to expected
hard-thermal-loop effective pole masses of approximately $gT$
\cite{Peshier:1995ty}. The mass acts as an effective IR regulator for
the scattering or ``emissions.''
	
In nature the initialization of the condensate is due to spontaneous
fluctuations. Because we choose a deterministic approach, Eq.\
(\ref{eq:diagrams2}) implies $\dot{n}_\c\sim n_\c$, i.e., condensation
does not occur, if $n_\c$ vanishes initially. To overcome this issue we
extract effective values for the chemical potential $\mu_\text{eff}$ and
the temperature $T_\text{eff}$ by fitting the IR region
($f_\text{IR}(p<m)$) of the distribution function to the Bose-Einstein
distribution function. If now $\mu_\text{eff}$ approaches $m$ (let us
name this point in time $t_{\text{onset}}$), we manually insert a finite
but negligibly small seed to the zero mode
$n_\c(t=t_\text{onset})=10^{-6}n_\text{init}$
\cite{PhysRevLett.74.3093,Xu:2014ega,MHZC16,PhysRevD.96.014020}. 

In the following simulations, $f_0=0.45 (2.0)$ has been chosen such that
the condensation criterion is generally fulfilled and vary in detail the
ratio $\sigma_{23}/\sigma_{22}$. Our simulations start in time at $t=0$
for solving Eqs.\ (\ref{eq:diagrams1}) and (\ref{eq:diagrams2}).
	
\section{Results}
	
\begin{figure}
  \centering
  \includegraphics[width=0.415\textwidth]{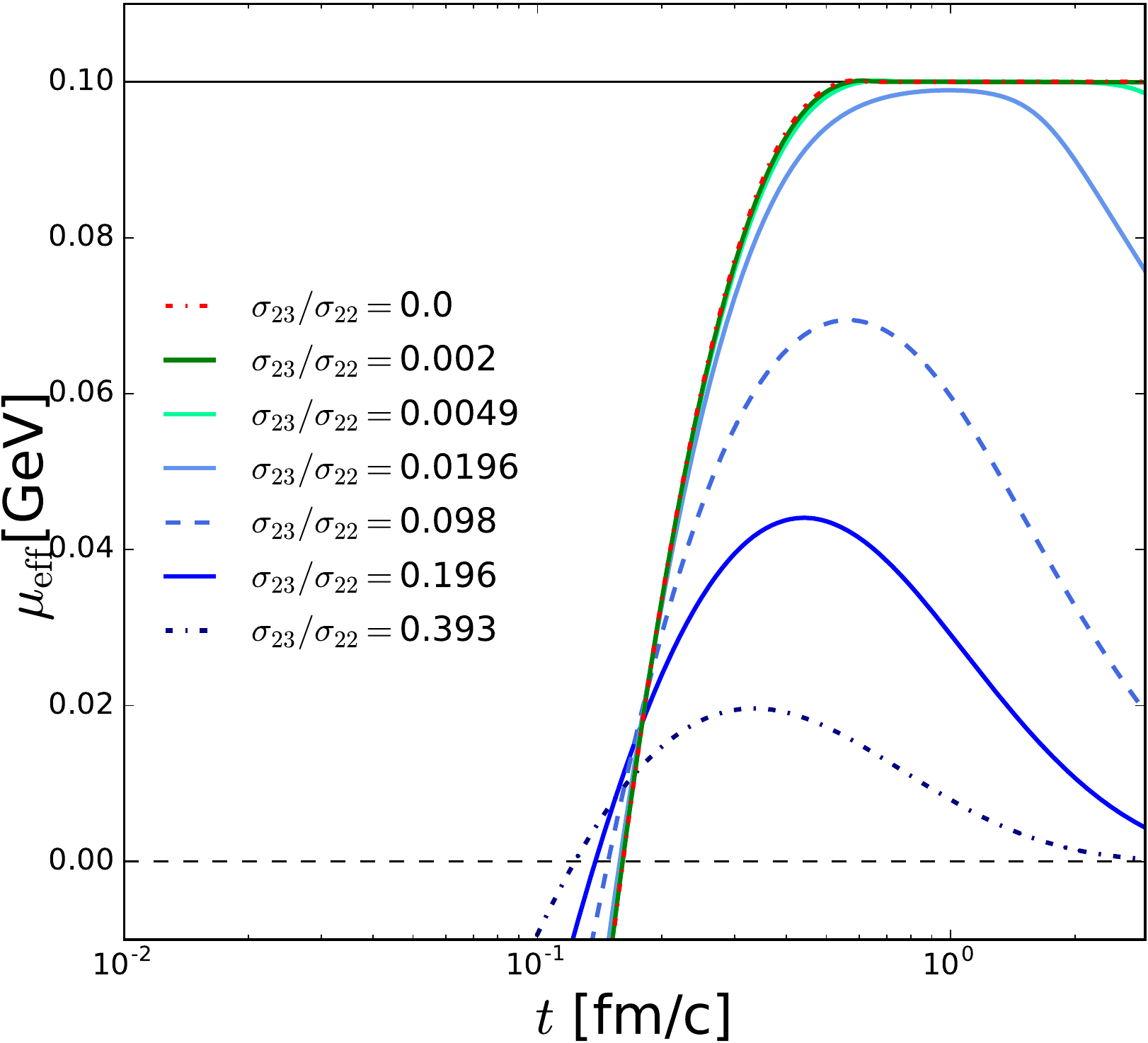}
  \caption{Time evolution of the effective chemical
    potential, with $m=100$ MeV at $f_0=0.45$. Bluish curves (inelastic)
    represents runs where condensation was not feasible, contrary to the
    greenish (inelastic) curves and red dashed line (elastic). The black
    line relates to the equilibrium condition for $2\leftrightarrow 2$
    processes ($\mu=m$) and the dashed line to inelastic processes
    ($\mu=0\text{ GeV}$). \label{fig:MainResult3}}
\end{figure}

In Figs.~\ref{fig:MainResult1_6plots} and \ref{fig:MainResult2}, the
main results are depicted and compared to the known case of the
evolution under solely binary scattering processes for several cross
section ratios $\sigma_{23}/\sigma_{22}$. The typical overpopulated
evolution for $2\leftrightarrow 2$ interactions consists of the particle
cascade toward the soft modes (Fig.~\ref{fig:MainResult1_6plots} (a))
followed by its decrease to the equilibrium distributions (e), while
generating a condensate until the equilibrium is reached.  The
introduction of $2\leftrightarrow 3$ kinetics, will dramatically change
this picture. The first observation is that the influx of particles
toward the soft modes (Fig.~\ref{fig:MainResult1_6plots} (b),
$\sigma_{23}/\sigma_{22}=0.0049$) is decelerated compared to the
previous case but still sufficient to hit the onset condition somewhat
later (Fig.~\ref{fig:MainResult3}), consequentially generating a
condensate. But once the Bose-Einstein shape for $\mu=m$ is recovered
[$t\simeq 0.9$ fm/c, Figs.\ \ref{fig:MainResult1_6plots} (f) and
\ref{fig:MainResult2}), we observe that the condensate decays, contrary
to the case considering only particle-number conserving
$2\leftrightarrow 2$ processes. For gradually larger values of
$\sigma_{23}/\sigma_{22}$, the characteristic particle transport towards
the soft modes is further damped and $\mu_\text{eff}$ never reaches the
onset of condensation. If $\sigma_{23}/\sigma_{22} \gtrsim 0.01$, then
no condensation into a BEC is observed.

The situation for the chemical potential $\mu_{\text{eff}}$ can be seen
in Fig.\ \ref{fig:MainResult3}. While for $\sigma_{23}=0$, an
equilibrium state with $\mu_{\text{eff}}=m=0.1\, \text{GeV}$ is reached,
this is only the case for the two smallest ratios
$\sigma_{23}/\sigma_{22}=0.002$ and $0.0049$, where $\mu_{\text{eff}}$
reaches $m$, but finally decreases again to reach the equilibrium state
with $\mu_{\text{eff}}=0$, as it is expected, if particle number is not
conserved.
	
\begin{figure*}[t!]  \centering
  \includegraphics[width=0.98\linewidth]{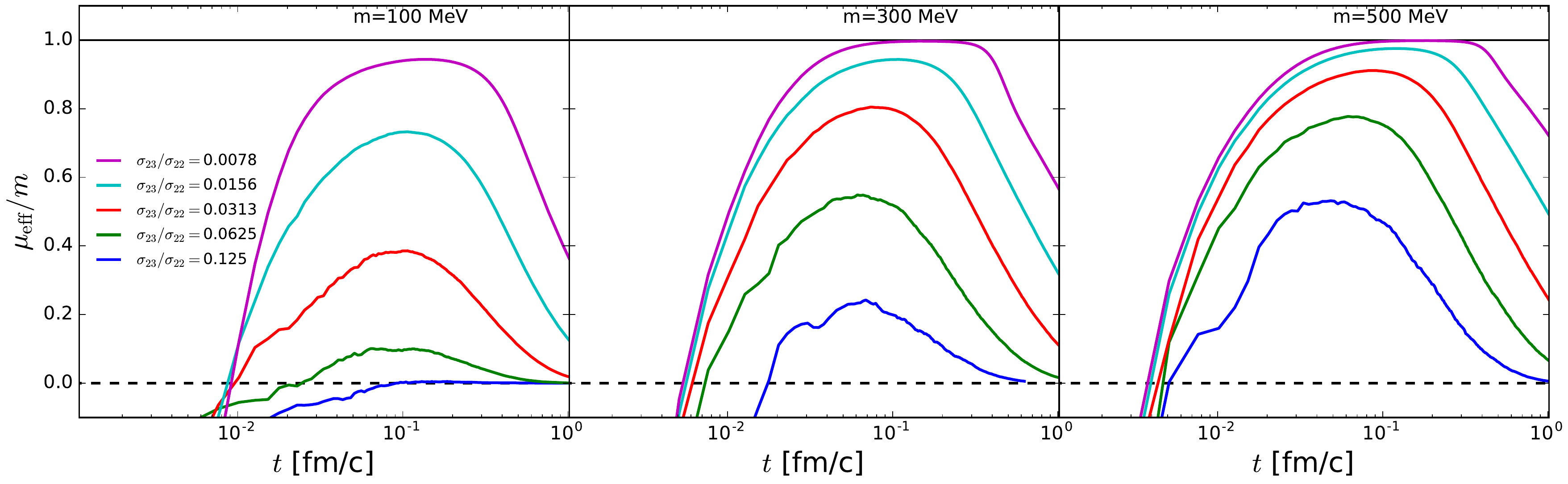}
  \caption{Time evolution of the effective chemical
    potential with respect to the masses at strongly overpopulated
    initial condition, $f_0=2$ for three different masses [100 (left),
    300 (middle), and 500 MeV (right)] and for different values of
    $\sigma_{23}/\sigma_{22}$. Solid black lines correspond to the
    equilibrium condition for $2\leftrightarrow 2$ processes, and the
    dashed lines correspond to inelastic processes. \label{fig:Mu_Masses}}
\end{figure*}

In Fig.\ \ref{fig:Mu_Masses} we show the time evolution of the effective
chemical potential, $\mu_{\text{eff}}$ in dependence of various masses,
$m=100$, 300 and 500 MeV. Please note for these calculations we employ a
strongly overpopulated initial condition with $f_0=2$. Inspecting the
calculations, only for $\sigma_{23}/\sigma_{22}=0.0078$ and for masses
$m=300$ and $500 \, \text{MeV}$ the chemical potential just touches the
mass limit $\mu_{\text{eff}} =m$, although no condensation will
start. The effect of earlier times for the onset of BEC formation of
heavier particles has also been found in a similar study with only
elastic collisions \cite{Blaizot:2015wga}.  Still, taking into account
inelastic $2 \leftrightarrow 3$ collisions, either for smaller or larger
masses no condensation occurs for the strongly overpopulated initial
condition. Only if $\sigma_{23}/\sigma_{22} \lesssim 0.005$ a momentary
and tiny BEC can develop.

\section{Conclusions}
	
In this paper we have investigated a complete Bose-Einstein condensation
of gluons within kinetic theory, explicitly including number changing
$2 \leftrightarrow 3$ processes.
In the presented scenario of an overpopulated nonequilibrium bosonic
system akin to Glasma-type initial conditions has been considered. The
bosons have been taken with a small but finite mass. The situation is
similar to the scenario of \cite{Tsutsui:2017uzd}. The cross sections
are not those of perturbative QCD. On the other hand binary scatterings
in thermal QCD are regulated by finite Debye-screening masses of order
$O(g T)$. Radiative perturbative QCD emissions are substantial for
describing the observed jet attenuation but also the significant
lowering of the shear-viscosity over entropy-density ratio
\cite{Xu:2007jv,Uphoff:2014cba}. The latter fact can be effectively
rephrased by significant $2\leftrightarrow 3$ isotropic collisions
\cite{El:2012cr}.

Our simulations have shown that a BEC may be formed for some limited
time if $\sigma_{23}/\sigma_{22}\ll1$. For present physical parameters
of the masses and overpopulation parameter, $f_0$, typically a BEC can
only appear if $\sigma_{23}$ is less than 1\% of $\sigma_{22}$. The
results suggest, that, as expected, particle-number conserving and
changing processes are counteracting mechanisms for the formation and
destruction of a BEC. We note that the individual collision integrals
scale with the occupation density of the system like $f^3$ (elastic) and
$f^4$ (inelastic), which resembles a sensitive scenario for possible
formation but also immediate decay of a BEC.

Summarizing, our calculations show that Bose-Einstein condensates in the
very early stage of heavy-ion collisions are highly unlikely, if
inelastic collisions are significantly participating in the dynamical
gluonic evolution.

\begin{acknowledgments}
  We are grateful to the LOEWE Center for Scientific Computing
  (LOEWE-CSC) at Frankfurt for providing computing resources.  We also
  acknowledge support by the Deutsche Forschungsgemeinschaft (DFG,
  German Research Foundation) through the Grant No.\ CRC-TR 211
  ``Strong-interaction matter under extreme conditions'' - Project No.\
  315477589-TRR 211. Z.X.'s work was financially supported by the
    National Natural Science Foundation of China under Grant
    No. 11575092. K.Z.'s work was financially supported by the BMBF under
    the ErUM-Data project and the AI research grant of SAMSON AG,
    Frankfurt.
\end{acknowledgments}

\bibliography{LiteraturBEC.bib}

\end{document}